\documentclass[11pt,a4paper]{article}
\usepackage{jcappub} 
\usepackage{amsmath,amssymb,graphicx}
\usepackage{xcolor}      
\usepackage{threeparttable,multirow,booktabs}
\usepackage{subcaption}
\usepackage{hyperref}

\newcommand{\lna}{\langle\ln A\rangle}

\title{Co-evolution of cosmic ray energy spectra, composition, and anisotropies}

\author[a,b]{Bing-Qiang Qiao}
\author[c,d]{Qiang Yuan}
\author[e,f,g]{Yi-Qing Guo}

\affiliation[a]{Deutsches Elektronen-Synchrotron (DESY), Platanenallee 6, 15738 Zeuthen, Germany}
\affiliation[b]{Institut f\"ur Physik und Astronomie, Universit\"at Potsdam, Golm Haus 28, 14476 Potsdam, Germany}
\affiliation[c]{Division of Dark Matter and Space Astronomy, Purple Mountain Observatory, Chinese Academy of Sciences, Nanjing 210023, China}
\affiliation[d]{School of Astronomy and Space Science, University of Science and Technology of China, Hefei 230026, China}
\affiliation[e]{State Key Laboratory of Particle Astrophysics, Institute of High Energy Physics, Chinese Academy of Sciences, Beijing 100049, China}
\affiliation[f]{University of Chinese Academy of Sciences, Beijing 100049, China}
\affiliation[g]{Tianfu Cosmic Ray Research Center, Chengdu, Sichuan 610213, China}

\emailAdd{yuanq@pmo.ac.cn}
\emailAdd{guoyq@ihep.ac.cn}

\abstract{
The origin of cosmic rays remains an unresolved fundamental problem in astrophysics. The synergy of multiple observational probes, including the energy spectra, the mass composition, and anisotropy is a viable way to jointly uncover this mystery. In this work, we propose that the energy-dependence of those observables in a wide energy range, from $O(10)$ GeV to ultrahigh energies of $10^{11}$ GeV, share quite a few correlated features, indicating a strong co-evolution which could be a consequence of the underlying origin of different source populations. We decipher these structures with a four-component model, i.e., the ensemble of Galactic sources, a local source close to the solar system, and the ensemble of two extragalactic source populations. In this scenario, the $O(10^2)$ GV hardening and $O(10)$ TV bump is due to the contribution of the local source, the knee is due to the maximum acceleration energy of protons by the Galactic source population, the second knee is due to the maximum acceleration energy of iron nuclei by Galactic sources, the dip feature between the two knees is due to the appearance of the extragalactic component, the ankle comes from the transition from one extragalactic component to the other, and the spectral suppression at the highest energies arises from the acceleration
limit of the second extragalactic component. The transition from Galactic to
extragalactic origin of cosmic rays occurs around $O(10^8)$ GeV, which is smaller than the ankle energy.
}

\begin{document}
\maketitle
\tableofcontents

\section{Introduction}
After more than one century of the discovery of cosmic rays (CRs), their origin 
is still unclear. The main challenge is the deflections of charged particles 
when they propagate in random magnetic fields everywhere in the Universe. 
It turns out that the arrival directions of CRs are almost isotropic, with 
small anisotropies (the amplitude is about $10^{-4}\sim10^{-3}$ below PeV
\cite{2017PrPNP..94..184A}), which erases the original locations of the sources. 
The other challenge is that the mass composition is not clear for indirect
measurements by ground-based experiments. Limited information with relatively
large uncertainty makes it difficult to fully address the problems about the 
origin, acceleration, propagation, and interaction of CRs. In recent years, 
with the developments of new generation space-borne and ground-based experiments,
significant progresses towards precise measurements of CRs have been achieved, 
which provides a very good opportunity to uncover the mystery of CR physics.
Particularly, a synergistic analysis of multi-messenger data, including spectra, 
anisotropies and mass composition, becomes feasible in light of the improved
precision. 

The joint discussion of multiple observables of CRs has been done in some works
\cite{2013FrPhy...8..748G,2013APh....50...33S,2016A&A...595A..33T,2019JCAP...03..017M,2022JCAP...07..006E,2023JCAP...05..024A,2024JCAP...01..022A,2025ApJ...979..225L,2024arXiv240313482Y}. Gaisser et al.
\cite{2013FrPhy...8..748G} proposed a parametric framework using three or four distinct 
astrophysical components to simultaneously describe the all-particle energy spectrum 
and mean logarithmic mass ($\lna$) of CRs. Sveshnikova et al. \cite{2013APh....50...33S}, approximately at the same time, demonstrated that nearby supernova remnants plus a distributed background can simultaneously explain the knee-region fine structure of the all-particle spectrum and the large-scale anisotropy, tying spectral features to anisotropy diagnostics. Thoudam et al.~\cite{2016A&A...595A..33T} argued that a single Galactic accelerator population cannot reproduce the observed spectrum and composition above \(\sim 2\times10^{16}\,\mathrm{eV}\), and therefore introduced a second Galactic component to extend the Galactic contribution up to \(\sim 10\,\mathrm{EeV}\). Their two-component Galactic scenario provides a unified description of the knee, the second knee, and the sub-ankle region without invoking an early extragalactic dominance. In contrast, Mollerach \& Roulet~\cite{2019JCAP...03..017M} reproduced the knee--to--second-knee structure by assuming rigidity-dependent cutoffs within a single Galactic component, together with the gradual emergence of an extragalactic flux at higher energies. Additionly, Eichmann et al. \cite{2022JCAP...07..006E} also indicated that radio-galaxy source populations can jointly reproduce the ultrahigh energy cosmic-ray (UHECR; $\gtrsim10^{18}$~eV) spectrum, composition, and large-scale anisotropies. Another two studies carried out by the Pierre Auger Collaboration also emphasized that a consistent interpretation of their observations requires going beyond a single homogeneous extragalactic source population, favoring scenarios with at least two distinct extragalactic contributions to simultaneously account for the observed spectrum, composition, and anisotropy features \cite{2023JCAP...05..024A,2024JCAP...01..022A}. Extending this methodology and based on 
LHAASO's unprecedented precision in measuring both the energy spectrum ($0.3-30$ PeV) 
and $\lna$ \cite{2024PhRvL.132m1002C}, Lv et al. \cite{2025ApJ...979..225L} and Yao et al. \cite{2024arXiv240313482Y} 
performed a comprehensive analysis spanning nine decades in energy from GeV to $10^{10}$ 
GeV. Their systematic investigation reveals compelling evidence for a non-negligible 
contribution from extragalactic CRs emerging at energies as low as $\sim10$ PeV, 
significantly below the traditional ankle feature ($\sim 5$ EeV) typically associated 
with the onset of extragalactic dominance. Anisotropies have not been extensively 
discussed in these works. The co-evolution of the spectra and anisotropies below 1 
PeV has been highlighted in 
Refs.~\cite{2013APh....50...33S,2019JCAP...10..010L,2019JCAP...12..007Q,2023ApJ...942...13Q}.
A nearby source at particular direction, with possible alignment effect along with the
local regular magnetic field, has been introduced to explain the spectral hardenings
at hundreds of GeV and softenings at tens of TeV, as well as the phase reversal of
the dipole anisotropies around 100 TeV. However, these studies did not cover the energy 
range above the knee region. 

Therefore, the all particle or individual spectra, composition, and anisotropy of CRs have their 
respective regions of precise measurement in observations due to the sensitivity of experimental 
instruments. When used independently to investigate the origin of CRs, regions where experimental
observations are not sensitive may fail to provide definitive results. However, by combining these 
messengers and complementing each other through common evolution, useful information can be obtained. Thus, this work aims to study the origin of CRs through the joint evolution of the total spectrum, component energy spectrum, composition, and anisotropy of CRs. The paper is organized as follows. Sec. II describes the observation data briefly. Sec. III presents our proposed four-component model, with the results given in Sec. IV. Lastly we conclude in Sec. V. 

\section{Observational data} 
The measurements of energy spectra for different species have big progresses recently,
revealing several new structures of the spectral shapes. A common hardening feature at a
rigidity of several hundred GV has been established by many direct detection experiments 
\cite{2007BRASP..71..494P,2010ApJ...714L..89A,2011Sci...332...69A,2015PhRvL.114q1103A, 2015PhRvL.115u1101A,2019SciA....5.3793A,2021PhRvL.126t1102A,2020PhRvL.125y1102A, 2022PhRvL.129j1102A,2023PhRvL.130q1002A}.
At slightly higher rigidity, $\sim 15$~TV, spectral softenings were further revealed by several 
experiments \cite{2017ApJ...839....5Y,2018JETPL.108....5A,2019SciA....5.3793A,2021PhRvL.126t1102A, 2022PhRvL.129j1102A,2022ApJ...940..107C,2023PhRvL.130q1002A,2022PhRvD.105f3021A,2025arXiv251105409D}.
Above 100 TeV (TV), hints of an additional spectral hardening were shown by a few experiments 
\cite{2024PhRvD.109l1101A,2024PhRvL.132e1002V}. This hardening is actually required to be 
reconciled with the all-particle spectra around PeV energies \cite{2020FrPhy..1524601Y}. 
For energies above $\sim100$ TeV, the measurements are mostly from ground-based indirect 
detection experiments. Several features have been observed, including the ``knee'' around 
$3\sim4$ PeV, the ``second knee'' around 200 PeV, the ``ankle'' around 5 EeV, and the 
suppression around 50 EeV 
\cite{1977ICRC....8..129D,1991ICRC....2...85F,1984JPhG...10.1295N, 1999APh....10..291G,2000A&A...359..682H,2005APh....24....1A,2008ApJ...678.1165A, 2013PhRvD..88d2004A,2008PhRvL.100j1101A,2008PhRvL.101f1101A,2013ApJ...768L...1A}. 
There is an additional hardening structure between the two knees, as reported by a few 
experiments \citep{2012APh....36..183A,2013PhRvD..88d2004A,2020APh...11702406B}.
Given more and more precise measurements, the overall spectra of CRs seem to be very 
complicated rather than structureless power-laws (see the top panel of Fig.~\ref{fig:aniso}). 
We can note that there are similarities among the spectral features at different energy bands. 
If we consider a hardening and the subsequent softening as one group, we can observe four 
such groups in the spectra from tens of GeV to the highest end: group 1 from 10 GeV to 10 TeV, 
group 2 from 100 TeV to 10 PeV, group 3 from 10 PeV to 1 EeV, and group 4 from 1 EeV to 100 EeV. 
Those similarities indicate that there are possible common mechanisms to form (at least some of) 
such structures.
 
The spectra of individual species are crucial to the understanding of the CR physics.
For energies below $\sim100$ TeV, direct detection experiments can well measure the 
individual spectra, which reveal the group 1 structures for several major species. 
At higher energies, however, it is usually difficult to distinguish different compositions 
with indirect detection technique, and the individual spectra have large (systematic) 
uncertainties. The measurements of average logarithmic mass number of particles, $\lna$, 
can in turn reflect the relative fractions of different species. Such measurements can be
done via measuring the lateral distribution of Cherenkov light \cite{2012NIMPA.692...98B},
the muon content in showers \cite{2005APh....24....1A,2013APh....42...15I,2024PhRvL.132m1002C}, 
or the longitudinal developments of showers 
\cite{2001ApJ...557..686A,2010PhRvL.104i1101A,2015APh....64...49A}.
See Ref. \cite{2012APh....35..660K} for an attempt to infer the average logarithmic mass 
from those observables. The results of $\lna$ in a wide energy range are shown in the 
second panel of Fig.~\ref{fig:aniso}. For energies below PeV, we directly calculate $\lna$ 
based on the individual spectra of the main species measured by ATIC \cite{2007BRASP..71..494P}, 
CREAM \cite{2010ApJ...714L..89A}, and AMS-02 \cite{2021PhR...894....1A}. 
With the increase of energy, the value of $\lna$ shows an increase from $\sim1.0$ at 10 
GeV to a plateau of $\sim 1.5$ at TeV-PeV, and increases again after PeV to reach a peak
value of $3.0\sim4.0$ around 100 PeV. Then it decreases to a valley of about 0.5 at about
1 EeV, after which it increases again. In general, the $\langle \ln A \rangle$ is inferred from the measurements of the depth of maximum development of air showers, $X_{\max}$, performed with fluorescence detectors. The relation between $X_{\max}$ and $\langle \ln A \rangle$ depends on the hadronic interaction models, such as SIBYLL, QGSJet, and EPOS, adopted to simulate CR interactions in the atmosphere. Nevertheless, as summarized in Ref. \cite{2013FrPhy...8..748G}, 
the overall energy dependence of $\langle \ln A \rangle$ is similar across models, 
showing consistent trends despite model-dependent shifts. 
In this work, we use the compilation of high-energy $\langle \ln A \rangle$ data provided in Ref.~\cite{2013FrPhy...8..748G}, in which the reconstruction is based on the SIBYLL hadronic interaction model. We do not separately consider different hadronic interaction models, since our analysis focuses on the common evolutionary features rather than the absolute normalization.
  
Anisotropies of arrival directions of CRs have been measured by ground-based air shower or 
underground muon detectors. It is obvious that for the energies less than $\sim$PeV, the 
amplitude of anisotropy does not exhibit a simple power-law dependence with energy. Instead, 
it reaches a maximum around 10 TeV and then decreases to a minimum near 100 TeV, forming a 
distinct ``trough''-like structure. Remarkably, the anisotropy phase simultaneously undergoes 
a reversal, from alignment with the IBEX-inferred local magnetic field direction to that of 
the Galactic center, resulting in a complementary ``reversal'' pattern 
\citep{2017PrPNP..94..184A,2014Sci...343..988S}. These fine-scale features are believed to 
be closely related to the influence of nearby sources, as suggested by recent theoretical
studies \citep{2019JCAP...10..010L, 2016PhRvL.117o1103A, 2019JCAP...12..007Q}. The KASCADE-Grande measurements of large-scale anisotropy did not yield statistically significant signals, with the maximum first-harmonic significance reaching only about $3.5\sigma$ in the $10^{15}$--$10^{16}\,\mathrm{eV}$ range. Owing to the large statistical and systematic uncertainties, the $99\%$ C.L.\ upper limits on the dipole amplitude of the order of $10^{-2}$ were reported \cite{2019ApJ...870...91A}. Of particular 
significance, the Pierre Auger Observatory also has performed a detailed analysis of the first-harmonic 
modulation in the right ascension (R.A.) distribution of CRs with energy more than several 
tens of PeV, determining both the dipole amplitude and phase in equatorial coordinates. 
For energies above 8 EeV, the measured dipole amplitude reaches $0.065^{+0.013}_{-0.009}$, 
with a statistical significance of $5.2\sigma$, and the corresponding right-ascension phase is observed at 
$100^\circ \pm 10^\circ$ \citep{2017Sci...357.1266P}. 
Furthermore, at lower energies ($E<8$~EeV), the reconstructed dipole components in each energy bin are not statistically significant and should therefore be regarded as consistent with upper limits, similar to the situation reported by KASCADE-Grande. Nevertheless, the East–West and Rayleigh measurements still hint at a possible trend: beginning around $E\sim 30$~PeV, the dipole amplitude appears to decrease toward $\sim 1$~EeV, before rising again and reaching the level of $\sim 10\%$ at $\sim 30$~EeV. Although this feature cannot be established at the current significance level, it is not excluded by the data and may be tested by future measurements with higher sensitivity. Meanwhile, the dipole phase shows a smooth transition: initially aligned with the Galactic-center direction, it gradually shifts toward $\alpha \sim 100^\circ$, with the transition occurring at a few~EeV \citep{2020ApJ...891..142A}.
This behavior strongly suggests an extragalactic origin for these CRs. 

It is very interesting to note that the structures of the spectra, composition, and
anisotropies show correlated evolution with energy. When there is spectral change 
(hardening or softening) on the energy spectra, some imprints on the $\lna$ and large-scale 
anisotropies (amplitude and phase) can be seen. For example, the softenings at $\sim15$ TV and $\sim3$ PeV in the spectrum correspond to local minima of $\lna$, while the hardenings around $100$ TeV and $30$ PeV correspond to local maxima of $\lna$. Moreover, the amplitude and phase of the large-scale anisotropies also show analogous variations at corresponding characteristic energies. Such a co-evolution of spectra, composition, and anisotropies give strong support of
a picture that CRs at different energy ranges have different origin. The sum of these
source populations, with different acceleration limits, source composition, and spatial
distributions, give rise to complicated but correlated structures of their spectra, 
$\lna$, and anisotropies.

\begin{figure*}[!htb]
\centering
\includegraphics[width=0.9\textwidth]{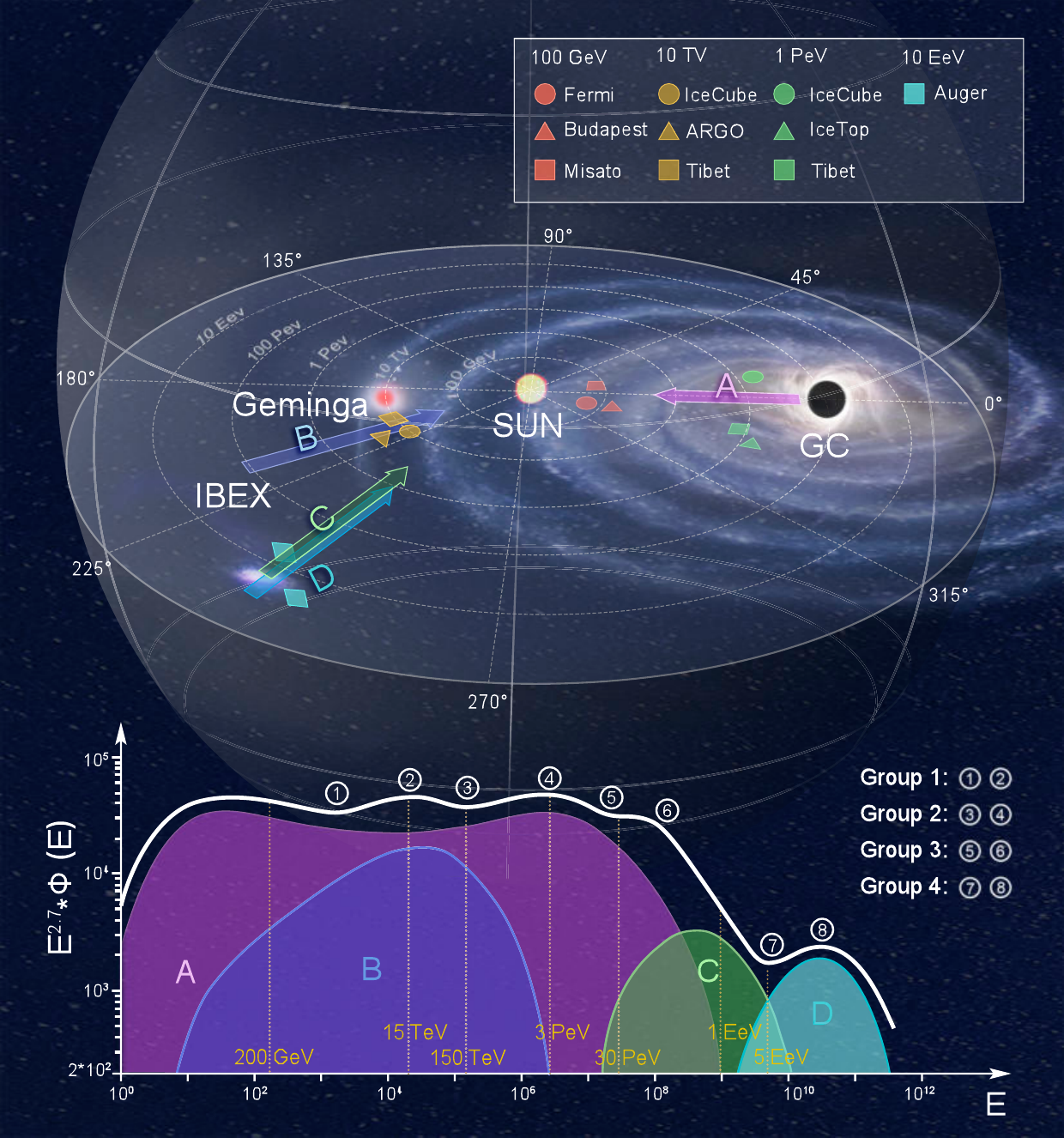}
\caption{Illustration of the contributions to CR spectra and arrival directions from
different source components. Filled circles, triangles, and squares show the measured
directions of anisotropies at different typical energies (red for 100 GeV, brown for
10 TeV, green for 1 PeV, and cyan for 10 EeV). It shows the evolution of the anisotropy
direction in the Galactic plane (Galactic coordinate) from the Galactic center (GC) at $\sim100$ GeV, to approximately the anti-GC 
(perhaps aligned with the local magnetic field as labelled by IBEX) around 10 TeV, to 
the GC again around PeV, and finally to the direction shown by green and cyan arrows 
around 10 EeV. The bottom part shows how the sum of these components result in complicated 
``ankle-knee'' structures in the spectra.}
\label{fig:cartoon}
\end{figure*}

\section{A four-component model}
In this work, we propose to understand the observed energy spectrum, composition and anisotropy 
of CRs with energies from $O(10)$ GeV to about $10^{20}$ eV using a four-component model, which 
includes the Galactic background sources (component A), one Galactic nearby source (component
B), and two extragalactic source populations (components C and D). Note that two extragalactic
source components are required here to fit the data of spectra and composition{\cite{2013FrPhy...8..748G,2019JCAP...03..017M,2014NuPhS.256..218S,2025ApJ...979..225L,2024arXiv240313482Y,2023JCAP...05..024A,2015PhRvD..92b1302G,2020PhRvD.101j3024M}.}
Fig.~\ref{fig:cartoon} is a cartoon plot to show the different source components mentioned above,
illustrating their contribution to the energy spectra and arrival directions (shown by arrows). 
The algebraic sum of these source components is responsible for the structures in the all-particle
spectra and the average logarithmic mass number, and the vector sum of these components gives 
energy evolution of the anisotropy amplitude and phase. 

\subsection{Galactic background sources}
Various types of sources in the Galaxy have been found to be able to accelerate CRs, such as
supernova remnants, massive star clusters, accretion disks and jets of black holes and so on.
While the detailed acceleration characteristics of these sources may be uncertain, 
phenomenologically a continuous source distribution with an average accelerated spectrum is
usually assumed, and the produced particles are injected in the Galaxy, experiencing a
diffusive transport process. The transport equation can be written as
\citep{2007ARNPS..57..285S} 
\begin{eqnarray}
\frac{\partial \psi_{\rm A}(\vec{r},p,t)}{\partial t} &=& q_{\rm A}(\vec{r}, p,t) + \nabla \cdot
\left( D_{xx} \nabla \psi - \vec{V_{c}}\psi \right) \nonumber\\
&+& \frac{\partial}{\partial p}\left[p^2D_{pp}\frac{\partial}{\partial p}\frac{1}{p^2}\psi \right]
- \frac{\partial}{\partial p}\left[ \dot{p}\psi - \frac{p}{3}
\left( \nabla \cdot \vec{V_c}\right) \psi \right] \nonumber\\
&-& \frac{\psi}{\tau_f} - \frac{\psi}{\tau_r},
\label{CRsPropagation}
\end{eqnarray}
where $\psi_{A}$ is the CR density per particle momentum interval at position 
$\vec{r}$, $q_{A}(\vec{r}, p, t)$ is the source function, $D_{xx}(\vec{r}, p)$ is the 
diffusion coefficient, $\vec{V_{c}}$ is the convection velocity, $D_{pp}(\vec{r}, p)$ 
is the diffusion coefficient in the momentum space which describes the reacceleration
of particles during the propagation, $\dot{p}$ is the momentum loss rate, $\tau_f$ and 
$\tau_r$ are the fragmentation and radioactive decaying time scales. The convection and
reacceleration are important for low energy CRs. For the energy range relevant for this
work ($E\gtrsim30$ GeV), their effects can be neglected. Usually a cylinder is adopted to 
describe the geometry of the propagation halo. Free escape is assumed at the border of 
the halo, namely $\psi_{A}(r, z, p) = \psi_{A}(r_h, z, p) = \psi_{A}(r, \pm z_h, p) = 0$.

The spatial distribution of the sources is parameterized as 
$f(r,z)=(r/r_\odot)^{1.25}\exp[-3.56(r-r_\odot)/r_\odot -|z|/(0.2~{\rm kpc})]$ 
\cite{2011ApJ...729..106T}, where $r_\odot=8.5$ kpc. The injection spectrum of CR nuclei are 
assumed to be an exponentially cutoff power-law function of particle rigidity ${\cal R}$ as 
$q_{\rm A}({\cal R}) = q_0^{\rm A} {\cal R}^{-\nu_{\rm A}} \exp[-{\cal R}/{{\cal R}_c^{\rm A}}]$, 
where $q_0^{\rm A}$ is the normalization factor, $\nu_{\rm A}$ is the spectral index, and 
${\cal R}_c^{\rm A}$ is the cutoff rigidity. Note that one or more breaks of the injection 
spectrum at low energies may be necessary to fit the wide-band data
\cite{2020ApJS..250...27B,2023RAA....23k5002P}. For the purpose of this work, we find that 
one single power-law with an exponentially cutoff is enough.

In the conventional model, the spatial diffusion coefficient is typically assumed to be
uniform everywhere throughout the Galaxy. However, it is natural to expect that the diffusion
coefficient should be spatially dependent, due to the non-uniform properties of the interstellar
medium (ISM). This picture is supported by recent observations of very-high-energy $\gamma$-ray
halos surrounding middle-aged pulsars \cite{2017Sci...358..911A,2021PhRvL.126x1103A}. 
It has been shown that the derived diffusion coefficient around these pulsars is smaller
by $10^2-10^3$ times than the average diffusion coefficient inferred from the CR
secondary-to-primary ratio (e.g., \cite{2017PhRvD..95h3007Y}). 
Consequently, we employ the spatially dependent propagation (SDP) scenario 
\cite{2012A&A...544A..16T,2016ApJ...819...54G} to describe the transport of particles.
The diffusion coefficient is smaller in a relatively thin disk (with vertical height 
$|z|\leq \xi z_h$), and bigger in the halo (with $|z|>\xi z_h$), where $z_h$ is the
height of the transport halo, $\xi$ is defined as the ratio between the thickness of the inner halo and that of the outer halo. A smooth connection of the diffusion coefficient between 
the disk and the halo is assumed. The detailed function form of the SDP part of the
diffusion coefficient can be found in \cite{2018PhRvD..97f3008G}.

At high energies (e.g., ${\cal R}>$PV), the gyro-radius of particles ($\gtrsim$pc) may be 
comparable to the coherent length of the local magnetic field, and the CR transport becomes 
less sensitive to small inhomogeneities of the ISM. In such a case, the diffusion coefficient
should be more uniform, and returns to the one in the halo. Therefore, the diffusion
coefficient is written as
\begin{equation}
D_{xx}(r,z, {\cal R} )= \left\{
\begin{array}{ll}
{D_{0}F(r,z)\beta^{\eta}\left(\frac{\cal R}
{{\cal R}_{0}} \right)^{\delta_{0}F(r,z)}}, & {\cal R}<1\,{\rm PV} \\
\\
D_{0}\beta^{\eta}\left(\frac{\cal R}
{{\cal R}_{0}} \right)^{\delta_{0}}, & {\cal R} > 1\,{\rm PV} \\
\end{array},
\right.
\label{eq:diffusion}
\end{equation}
where $\beta$ is the particle's velocity in unit of the light speed, $D_0$ and $\delta_0$ 
are constants characterizing the diffusion coefficient and its rigidity dependence in the 
halo, $\eta$ is a phenomenological constant in order to fit the low-energy data, and $F(r,z)$ 
describes the spatial variation of the diffusion coefficient \cite{2018PhRvD..97f3008G}. 



In this work, we adopt the diffusion re-acceleration model, with the diffusive re-acceleration 
coefficient $D_{pp}$, which correlates with $D_{xx}$ via 
$D_{pp}D_{xx} = \frac{4p^{2}V_{A}^{2}}{3\delta(4-\delta^{2})(4-\delta)}$, where $V_A$ 
is the Alfv\'en velocity, $p$ is the momentum, and $\delta$ is the rigidity dependence slope of 
$D_{xx}$ \cite{1994ApJ...431..705S}. We use the DRAGON code to solve the transport equation
\cite{2017JCAP...02..015E}. For energies smaller than tens of GeV, the fluxes of CRs are 
further suppressed by the solar modulation effect, for which we use the force-field 
approximation \cite{1968ApJ...154.1011G}. The main propagation parameters are:
$D_0=8.1\times10^{28}$ cm$^2$~s$^{-1}$, $\delta_0=0.56$, $\eta=0.05$, $N_m=0.9$, $\xi=0.09$, 
$n=4$, $V_A=6.0$ km~s$^{-1}$, $z_h=4.5$ kpc. The reference rigidity is ${\cal R}_0\equiv1$ GV.
We find that a unified spectral index of $\nu_{\rm A}=2.41$ and cutoff rigidity of
${\cal R}_c^{\rm A}=8$ PV can well fit the spectra of individual species and all particles, 
and the average logarithmic mass. The normalization parameter of each species can be found 
in Table~\ref{tab:para_inj}.

\begin{table*}
\begin{center}
\caption{Normalization parameters of the four source components.}
\renewcommand{\arraystretch}{1.15}
\setlength{\tabcolsep}{8pt}
\begin{tabular}{|c|c|c|c|c|}
\hline
& Component A & Component B & Component C & Component D \\
\hline
Element 
& $q_0^{\rm A}$$^\dagger$ 
& $q_0^{\rm B}$ 
& $q_0^{\rm C}$$^*$ 
& $q_0^{\rm D}$$^*$ \\
\hline
& $[({\rm m}^2\!\cdot\!{\rm sr}\!\cdot\!{\rm s}\!\cdot\!{\rm GeV})^{-1}]$ 
& [GeV$^{-1}$] 
& $[({\rm m}^2\!\cdot\!{\rm sr}\!\cdot\!{\rm s}\!\cdot\!{\rm GeV})^{-1}]$ 
& $[({\rm m}^2\!\cdot\!{\rm sr}\!\cdot\!{\rm s}\!\cdot\!{\rm GeV})^{-1}]$ \\
\hline
p  & $2.83\times 10^{-2}$ & $1.8\times 10^{52}$ & $5.0\times 10^{1}$   & $6.8\times 10^{-5}$ \\
He & $1.29\times 10^{-3}$ & $1.5\times 10^{52}$ & $6.0\times 10^{-1}$  & $2.5\times 10^{-5}$ \\
C  & $4.47\times 10^{-5}$ & $4.6\times 10^{50}$ & $1.0\times 10^{-1}$  & $4.0\times 10^{-6}$ \\
N  & $7.53\times 10^{-6}$ & $4.5\times 10^{49}$ & $3.0\times 10^{-2}$  & -- \\
O  & $5.43\times 10^{-5}$ & $4.7\times 10^{50}$ & $6.0\times 10^{-2}$  & $2.0\times 10^{-6}$ \\
Ne & $1.14\times 10^{-5}$ & $8.0\times 10^{49}$ & $4.0\times 10^{-2}$  & -- \\
Mg & $1.42\times 10^{-5}$ & $8.0\times 10^{49}$ & $2.0\times 10^{-2}$  & -- \\
Si & $1.09\times 10^{-5}$ & $7.5\times 10^{49}$ & $1.0\times 10^{-2}$  & -- \\
Fe & $1.02\times 10^{-5}$ & $3.5\times 10^{49}$ & $1.0\times 10^{-3}$  & $1.0\times 10^{-7}$ \\
\hline
\end{tabular}\\[0.4em]

{\footnotesize
$^\dagger$ The normalization is set at kinetic energy per nucleon $E_k = 100$ GeV/n.\\
$^*$ The normalization is set at particle's total kinetic energy $E = 1$ GeV.
}
\label{tab:para_inj}
\end{center}
\end{table*}

\subsection{Galactic nearby source}
Local source(s) has been proposed as one possible origin of the observed features of 
energy spectra ($\sim200$ GV hardenings and $\sim15$ TV softenings) and dipole anisotropies
\cite{2013APh....50...33S,2019JCAP...10..010L,2019JCAP...12..007Q,2025arXiv251106733Y}. Among the observed 
candidate sources such as supernova remnants, it has been shown that, under the framework
of slow diffusion in the Galactic disk, Geminga could be the right source to have proper
contribution to the locally observed CR spectra \cite{1994APh.....2..257J,2022ApJ...930...82L}. The direction
of Geminga, (R.A., Decl.)~$=(6^h34^m,\,17^{\circ}46')$, is also consistent with the direction 
of the dipole anisotropy phase below 100 TeV\footnote{Considering the alignment effect along 
the local regular magnetic field \cite{2014Sci...343..988S}, other sources located in the outer Galaxy hemisphere might also reproduce the measured dipole anisotropy \cite{2016PhRvL.117o1103A}.}
The electrons and positrons produced by the pulsar wind nebula associated with the local
source can further explain the positron excess and the total electron plus positron
spectra \cite{2022PhRvD.105b3002Z}. Here we also take Geminga as an illustration of local
source. Its distance is adopted as 250 pc, and age is adopted as $3.4\times10^5$ yr
\cite{2005AJ....129.1993M}.

The propagation of nuclei from the nearby source can be calculated using the Green's 
function method, assuming a spherical geometry with infinite boundary conditions. 
Assuming instantaneous injection from a point source, the CR density as a function of 
space, energy, and time can be calculated as
\begin{equation}
\psi_{\rm B}(r, E, t) = \dfrac{q_{\rm B}(E)}{\left(\sqrt{2\pi} \sigma\right)^3} 
\exp \left(-\dfrac{r^2}{2\sigma^2} \right)~,
\end{equation}
where $q_{\rm B}(E)$ is the injection spectrum as a function of rigidity, 
$\sigma(E, t)=\sqrt{2D(E)t}$ is the effective diffusion length within 
time $t$. The diffusion coefficient $D(E)$ takes the solar system value of 
Eq.~(\ref{eq:diffusion}). The function form of $q_{\rm B}(E)$ is assumed to 
be power-law with an exponential cutoff, i.e., $q_{\rm B}(E)=q_0^{\rm B}E^{-\nu_{\rm B}} \exp(-E/(Z{\cal R}_c^{\rm B}))$. Through fitting to the
data, we find that the spectral index $\nu_{\rm B}$ is about 2.32, and the cutoff 
rigidity is about 40 TV for all species. The normalization parameter $q_0^{\rm B}$ 
of different species is given in Table~\ref{tab:para_inj}. 

\subsection{Extragalactic low- and high-energy sources}
With the increase of energy, it is expected that CRs should experience a transition from
the Galactic origin to the extragalactic origin. The acceleration end of Galactic sources
and the transition energy is unclear yet. Different models are proposed to explain the ankle 
and cutoff structures of ultrahigh energy CRs \cite{2008PhRvL.100j1101A,2008PhRvL.101f1101A}, 
which predict different transition energy. In one type of models, the highest energy 
suppression of the spectrum is due to the GZK process, i.e., interaction between protons 
and the cosmic microwave background (CMB) photons \cite{2006PhRvD..74d3005B}. In this
scenario the ankle is explained by the pair production of electrons and positrons due to 
the $p\gamma$ interaction. This model requires that the composition of ultrahigh energy 
CRs is dominated by protons, which seems to be different from the recent measurements 
\cite{2019ICRC...36..482Y}. The other type of models suggest the highest energy cutoff 
is due to the acceleration limit of extragalactic sources and the ankle is ascribed to 
the transition from Galactic origin to extragalactic origin 
\cite{2011APh....34..620A,2012APh....39..129A}. In this case, the extragalactic component appears at energies higher than several EeV, and there will be a gap between 0.1 
and 1 EeV if one assumes that the knee is due to the acceleration limit of protons from 
Galactic sources. As noted earlier, the idea of employing two extragalactic components to account for this gap was already explored in early studies \cite{2015PhRvD..92b1302G,2020PhRvD.101j3024M}. Hillas further suggested that the gap could be bridged by introducing a second Galactic component (Hillas Component B) \cite{2005JPhG...31R..95H}.

In this work, we basically follow the latter type of models, and employ two extragalactic components, denoted here as C and D, to reproduce the multimessenger data.
The two extragalactic components were also found to be necessary to explain the measured mass 
composition \cite{2013FrPhy...8..748G,2025ApJ...979..225L,2024arXiv240313482Y}. The spectral form for the 
extragalactic components is parameterized as \cite{2019JCAP...03..017M}
\begin{equation}
\psi_i(E) = q_0^i E^{-\nu_i} \times \exp \left(-\frac{E}{Z R_c^i}\right) \times 
\frac{1}{\cosh \left[\left(\frac{Z R_s}{E}\right)^{\beta}\right]},
\end{equation}
where $i$ denotes the low-energy (component C) and high-energy (component D), $q_0^i$ 
is the flux normalization, $\nu_i$ is the spectral index, $Z$ is the particle charge, 
$R_c^i$ is the cutoff rigidity. The adopted spectral parameters are $\nu_{\rm C}=2.45$, ${\cal R}_c^{\rm C}=1.0$ EV and
$\nu_{\rm C}=1.95$, ${\cal R}_c^{\rm D}=15$ EV. The normalization
parameters are given in Table~\ref{tab:para_inj}. 
It should be noted that the effects of propagation are not considered here, and therefore the parameters correspond to the values observed at Earth.
The cosh term mainly describes the suppression of the low-energy spectrum due to the magnetic horizon effect from extragalactic magnetic fields \cite{2005PhRvD..71h3007L,2006ApJ...643....8B,2013JCAP...10..013M} or the shielding effect of Galactic
magnetic field to extragalactic CRs, with characteristic shielding rigidity $R_s=60$ PeV, which is a free parameter determined through reproducing the observed spectrum, composition,
and anisotropy features,
and finally $\beta=1.6$ is a parameter to describe the smoothness of the low-energy 
shielding. 


In order to reproduce the energy dependence of the anisotropy of extragalactic cosmic rays, we adopt the same functional form of the diffusion coefficient as in Eq.~(\ref{eq:diffusion}), but with different parameters from the Galactic case. Given the large uncertainties in intergalactic magnetic fields and the absence of a well-established diffusion model, the diffusion coefficient is treated as an effective quantity that characterizes the energy dependence of particle transport and thus governs the evolution of the dipole anisotropy. Observationally, the anisotropy above EeV energies exhibits a stronger energy dependence than that at PeV energies. To account for this behavior, we determine the energy-dependence index $\delta_0$ and normalization $D_0$ by fitting the anisotropy data above EeV energies. Note that, these parameters should be regarded as phenomenological quantities, rather than as direct representations of the underlying physical diffusion properties.

\begin{figure*}[!htbp]
\centering
\includegraphics[width=0.78\textwidth]{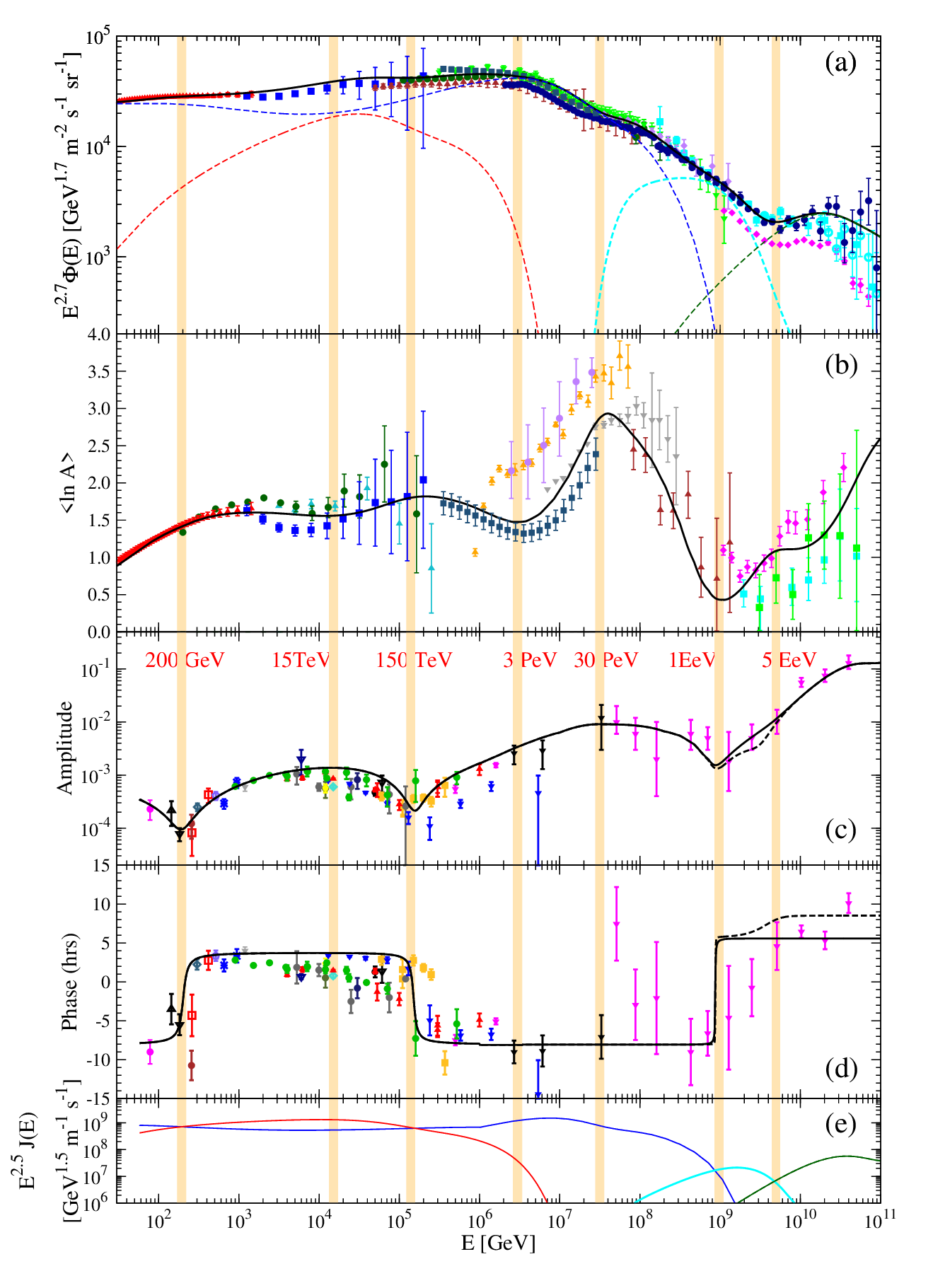}
\caption{The co-evolution of the all-particle spectrum (panel a), the mean logarithmic mass 
$\lna$ (panel b), the amplitude (panel c) and phase (panel d) of the equatorial dipole component of the large scale anisotropies, and the strength of CR streaming (defined as \( J(E) = \lvert D(E) \nabla \phi \rvert \); panel e), for energies from tens of GeV to $\sim$100 EeV. Black solid lines show the total model calculation results. In panels a and e, the contributions from components A (blue), B (red), C (cyan), and D (green) are also shown. The dashed line in panel d shows the expected phase for a different direction of the extragalactic component D from that of component C. The vertical bands label the characteristic energies of correlated features of the multi-messenger observables. 
References of the measurements are:
energy spectra 
\cite{2021PhRvL.127b1101A,2021PhRvL.126d1104A,2020PhRvL.124u1102A,2018PhRvL.121e1103A,
2017PhRvL.119y1101A,2015PhRvL.115u1101A,2015PhRvL.114q1103A,2017ApJ...839....5Y,
2009ApJ...707..593A,2003APh....19..193H,2008ApJ...678.1165A,2019PhRvD.100h2002A,
2013APh....42...15I,2020PhRvD.102l2001A,2020PhRvL.125l1106A,2020PhRvD.102f2005A,
2008PhRvL.100j1101A,2012NIMPA.692...98B,2020APh...11702406B,2016APh....80..131A,
2018ApJ...865...74A,2024PhRvL.132m1002C};
mean logarithmic mass \cite{2022AdSpR..70.2696T,2009BRASP..73..564P,2005APh....24....1A,
2013APh....47...54A,2013NIMPA.700..188A,2019PhRvD.100h2002A,2012NIMPA.692...98B,
2012APh....35..660K,2013JCAP...02..026P,2021ApJ...909..178A,2024PhRvL.132m1002C});
anisotropies \cite{1973ICRC....2.1058S,1975ICRC....2..586G,1981ICRC...10..246B,
1981ICRC....2..146A,1983ICRC....3..383T,1985P&SS...33..395N,1985P&SS...33.1069S,
1987ICRC....2...22A,1989NCimC..12..695N,1995ICRC....4..639M,1995ICRC....4..648M,
1995ICRC....4..635F,1995ICRC....2..800A,1996ApJ...470..501A,1997PhRvD..56...23M,
2003PhRvD..67d2002A,2005ApJ...626L..29A,2007PhRvD..75f2003G,2009ApJ...692L.130A,
2009NuPhS.196..179A,2009ApJ...698.2121A,2010ApJ...718L.194A,2012ApJ...746...33A,
2013ApJ...765...55A,2015ApJ...809...90B,2017ApJ...836..153A,2018ApJ...861...93B,
2019ApJ...883...33A,2024ApJ...976...48A,2017Sci...357.1266P,2020ApJ...891..142A}.
Note that for energies $\lesssim100$ TeV, the total spectra and mean logarithmic mass
are derived from direct detection experiments such as AMS-02, CREAM, and NUCLEON.
}
\label{fig:aniso}
\end{figure*}

The relevant parameters of components C and D are summarized in Table~\ref{tab:para_inj} and are obtained by manual adjustment here. We restrict the spectral 
shape parameters to be the same for all species of each source component and leave only their 
normalizations free to be tuned. This constraint reduces significantly the degeneracy of parameters.
However, due to the lack of measurements of individual spectra at high energies, additional 
uncertainties of the abundance parameters may still exist. We expect that, within reasonable 
ranges of the parameters, our main conclusions, as shown in Fig.~\ref{fig:aniso}, remain unaffected.

Since these parameters need to simultaneously reproduce multiple cosmic-ray observables, they are not arbitrarily varying. Our aim was to identify, based on previous experience, a consistent set of model parameters that enables to match the energy spectrum, $\lna$, and anisotropy data. The uncertainties of the model parameters arise from both experimental errors and parameter degeneracy, and within their reasonable ranges our main conclusions, as shown in Fig.~\ref{fig:aniso}, remain robust.

\begin{figure*}[!htb]
\centering
\includegraphics[width=0.9\textwidth]{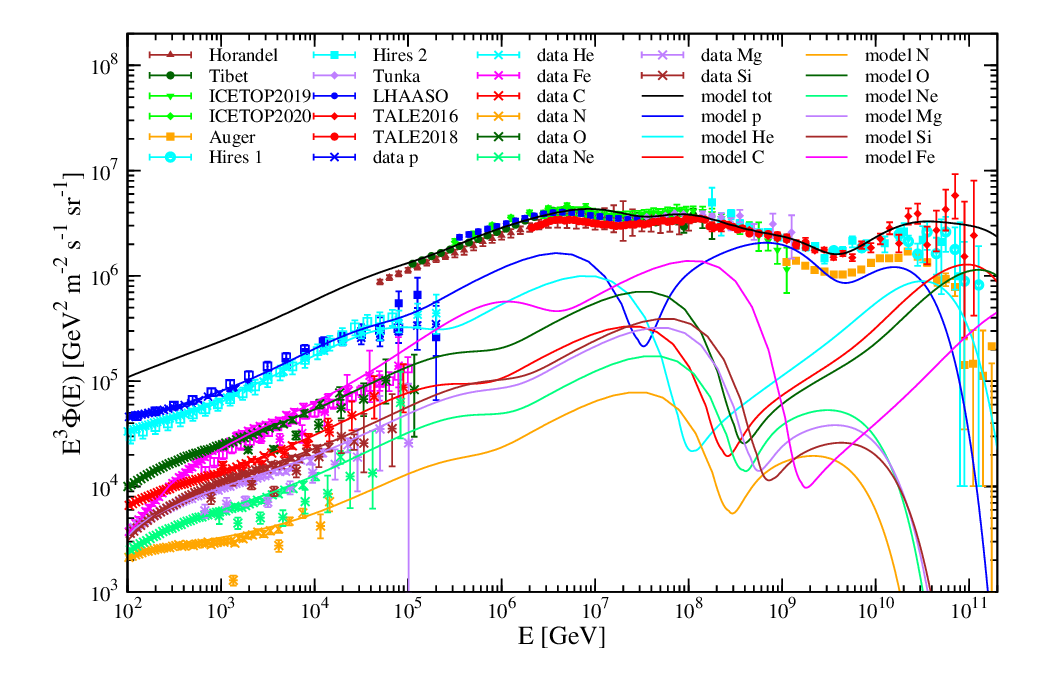}
\caption{Model calculated spectra of all particles and several major species (p, He, C, N, O, Ne, Mg, Si, Fe), compared with the measurements. References of the all-particle spectrum are the same as in Fig.~\ref{fig:aniso}, and references for individual mass groups are: protons \citep{2015PhRvL.114q1103A,2017ApJ...839....5Y,2022AdSpR..70.2696T,2019SciA....5.3793A}, He \citep{2015PhRvL.115u1101A,2017PhRvL.119y1101A,2017ApJ...839....5Y,2021PhRvL.126t1102A}, C \citep{2017PhRvL.119y1101A,2009ApJ...707..593A}, N \citep{2018PhRvL.121e1103A,2009ApJ...707..593A}, O \citep{2017PhRvL.119y1101A,2009ApJ...707..593A}, Ne \citep{2020PhRvL.124u1102A,2009ApJ...707..593A}, Mg \citep{2020PhRvL.124u1102A,2009ApJ...707..593A}, Si \citep{2020PhRvL.124u1102A,2009ApJ...707..593A}, Fe \citep{2021PhRvL.126d1104A,2009ApJ...707..593A,2021PhRvL.126x1101A}.
}
\label{fig:spec}
\end{figure*}

\section{Results}

The results of the model fittings together with the measurements for the energy spectra,
$\lna$, amplitudes and phases of the dipole anisotropies are shown in Figs. \ref{fig:aniso}
and \ref{fig:spec}. Below we discuss each of these observables. 

\subsection{Spectra of all-particle and individual composition}
Panel (a) of Fig. \ref{fig:aniso} shows the all-particle energy spectrum expected from the
four source components (blue for A, red for B, cyan for C, and green for D) and their sum
(black solid line). Note that, for the low-energy part below $\sim100$ TeV, the spectrum 
and $\lna$ are derived according to the direct measurements of major compositions such as 
protons, He, CNO, NeMgSi, and Fe. The main features of the spectrum, e.g., the four groups 
``hardening-softening'' structures, can be properly reproduced by the sum of different 
source components, reflecting complicated origin of CRs over the entire energy range.
The group 1 ``hardening-softening'' feature between 10 GeV and 100 TeV comes mainly from
the superposition of Galactic components A and B. Due to the sum of all compositions,
the actual hardening energy of the all-particle spectrum is about TeV. With the decrease 
of contribution from component B above tens of TeV, the recovery of component A dominance 
results in the group 2 feature, producing the knee with its acceleration limit. 
The appearance of the extragalactic component C results in a hardening around 30 PeV, 
and the acceleration limit of iron from component A results in the second knee around 
200 PeV \cite{2002JHEP...12..032C}. These two effects form the group 3 feature. Finally, the transition between 
extragalactic component C and D gives rise to the ankle around 5 EeV, and the acceleration 
limit of component D gives the highest energy cutoff (group 4). 

According to current observations, there remains a certain level of inconsistency among the all-particle spectra measured by different ground-based experiments in the energy range of 1 EeV–100 EeV, with the Pierre Auger reporting significantly lower fluxes than HiRes and Telescope Array (TA) \cite{2020APh...11702406B,2016APh....80..131A,
2018ApJ...865...74A}. Nevertheless, all the experiments reveal similar spectral features at the same characteristic energies. The focus of our work is to investigate whether the fine structures embedded in the spectra of different messengers exhibiting a common evolutionary trend with energy. Consequently, our model expectations are not tailored to any single experiment but are instead intended to capture the characteristic structures observed by multiple experiments. In order to remain consistent with the nuclear species considered in component C as far as possible, we have also included substantial contributions from heavy nuclei (C, O, Fe) in component D. This results in a less pronounced cutoff near ~50 EeV. However it does not affect the main conclusions in our study.

\subsection{Mean logarithmic mass}
The mean logarithmic mass is another critical messenger carrying information about the
composition of CRs. It is very interesting to see that the spectral structures due to
different source components also leave corresponding imprints on the mean logarithmic mass,
primarily due to the basic assumption of this work, that the spectra of different 
compositions have a charge-dependence. Therefore, whenever a new source component starts
to appear, the mean mass composition varies from heavy to light since the new component
is always proton dominated at first. When the contribution from the new component starts 
to decrease, the composition becomes heavy again. Then we see that each group of 
hardening-softening spectral feature corresponds to a ``bump-dip'' structure of $\lna$, 
as labelled by vertical bands in Fig. \ref{fig:aniso}. Such a correlation has been
highlighted by the recent high-precision measurements of the all-particle spectrum
and $\lna$ with LHAASO \cite{2024PhRvL.132m1002C}. Although relatively large uncertainties
of the measurements of $\lna$ exist in other energy bands, the correlation between
$\lna$ and the spectra is visible.

\subsection{Large scale anisotropy}
The dipole component of the large scale anisotropy reflects the streaming of CRs. The measured
amplitudes and phases of the dipole anisotropy show complicated energy evolution, which may
reflect the sum effect of multiple streamings. As shown in panels (c) and (d) of 
Fig.~\ref{fig:aniso}, the characteristic energies of the anisotropy features also show good 
correspondence with those of spectra and $\lna$. The dipole anisotropy can be well understood 
in the four component source model. 

For source component A, the CR streaming points from the GC to anti-GC. The nearby source 
(component B) generates a different streaming with direction pointing from the source to the 
Earth. The sum of the two streamings then gives the observed anisotropy evolution. At $\sim150$ 
TeV a phase reversal is expected due to the nearby source streaming becoming sub-dominant 
compared with the Galactic background sources.
In contrast, at energies of around 100 GeV the contribution from nearby source gradually becomes dominant, leading naturally to another phase reversal; this indicates that the phase shifts from the GC direction toward that of the nearby source. 
The new observations of proton anisotropy by Fermi space telescope seems to show such 
a trend \cite{2019ApJ...883...33A}. However, the significance is relatively low due to 
limited statistics. Future measurements by space-borne detector such as HERD 
\cite{2014SPIE.9144E..0XZ} and VLAST \cite{2024NuScT..35..149P} are able to test this 
expected anisotropy feature. The local regular magnetic field may further regulate the
directions of low-energy CR streamings, resulting in changes of detailed numbers of the
phase \cite{2014Sci...343..988S,2016PhRvL.117o1103A}. The basic picture discussed in this
work still holds, and we do not focus on such details.

From $\sim150$ TeV up to the end of the Galactic contribution, the main contribution to 
the anisotropy is component A, with phase of the GC. From $\sim 30$ PeV, component C starts 
to take effect, and we can see a break of the energy-dependence of the amplitude. The phase
keeps unchanged until $\sim1$ EeV, at which the streaming from source component C becomes 
dominant. 
This transition is expected to be relatively sharp rather than smooth, because the dipole phase is determined by the direction of the total streaming vector, $\mathbf{J}_{\rm tot}=\sum_i\mathbf{J}_i$. Around $\sim 1$ EeV, the dominant contribution to $\mathbf{J}_{\rm tot}$ changes from component A to component C. Since these two components have significantly different streaming directions, the direction of $\mathbf{J}_{\rm tot}$ changes rapidly when $|\mathbf{J}_A|\sim|\mathbf{J}_C|$, producing the sharp phase flip near $\sim 1$ EeV. A similar sharp transition around $\sim 100$ TeV provides another example of this component-switching behaviour. Moreover, due to the limited knowledge about the propagation of CRs in the extragalactic space, we adopt a phenomenological approach to calculate the anisotropy of 
extragalactic sources via simply introducing a streaming with given direction, and adjusting 
its strength and energy-dependent slope to fit the measurements. 
Around 5 EeV, component D further dominates over component C, and additional changes may
exist if the phases of components C and D are different. 
Here, we consider two specific cases. In the baseline configuration, both components are assigned the same direction, $(l,b)_{\rm C,D} = (251^\circ,-32^\circ)$, which reproduces the observed phase behaviour. In an alternative configuration, the two components are assigned different directions, $(l,b)_{\rm C} = (251^\circ,-32^\circ)$ and $(l,b)_{\rm D} = (155^\circ,-32^\circ)$, to explore the impact of the extragalactic geometry. The corresponding results are illustrated by the solid and dashed lines in panel (d) of Fig.~\ref{fig:aniso}.

\section{Conclusion and discussion}
Through a detailed investigation of the multi-messenger observables of CRs, we notice that 
in the very wide energy range from tens of GeV to highest end of the CR spectra, the energy 
spectrum, average logarithmic mass, and large-scale anisotropy co-evolve with energy. 
Such a co-evolution indicates that these features share common origins. 
We then propose a four-component model, with two Galactic source components and two
extragalactic source components, to account for the wide-band observational features. 
For each component, we assume different compositions share similar rigidity spectrum and
adjust their abundance to fit the measurements. The observed features of the spectrum,
$\lna$, and dipole anisotropy amplitude and phase, can be reproduced as the relative
weights of different components vary with energy. Particularly, the spectrum is the algebraic 
sum of the four components, and the anisotropy is the vector sum of the four components. 

The structures of the energy spectrum can be classified into 4 groups of ``hardening-softening''
features. From low to high energy, they are group 1 with $O(10^2)$ GV hardening and $O(10)$ 
TV softening, group 2 with $O(10^2)$ TeV hardening and $\sim3$ PeV softening (also known as
the ``knee''), group 3 with $\sim30$ PeV hardening and $\sim200$ PeV softening (the second
``knee''), and group 4 with $\sim 5$ EeV hardening (the ``ankle'') and the $\sim50$ EeV
suppression. Correspondingly there are imprints on $\lna$ and dipole
anisotropy amplitude and phase at these characteristic energies. The group 1 feature is
interpreted as the sum of Galactic source components A and B. The CR streaming from
component B dominates over component A between $\sim100$ GeV and $100$ TeV, resulting in
two phase reversals of the dipole anisotropy. With the decrease of the contribution from
component B for energy above tens of TeV, the recovery of component A and its acceleration
limit give the group 2 feature. The appearance of extragalactic component C and the decrease
of component A gives the group 3 feature, and finally the transition from component C to
component D and the acceleration limit of component D produce the group 4 feature. 

While such a four component scenario can account for the complicated structures on the
energy spectrum, $\lna$, and dipole anisotropy, we note that the current measurements
have relatively large uncertainties, especially for the composition and anisotropy
measurements above tens of PeV. Measurements of the mass composition between TeV and
100 TeV, as well as the anisotropy below TeV, relevant to direct detection experiments, 
are also uncertain. Improved measurements of these quantities in both low and high energy 
bands by future direct and indirect detection experiments are very crucial in further 
constraining the source populations of CRs. 

\acknowledgments
This work is supported by the National Key R\&D program of China (No. 2024YFA1611402), the 
National Natural Science Foundation of China (Nos. 12220101003, 12333006, 12275279), and 
the Project for Young Scientists in Basic Research of Chinese Academy of Sciences (No. YSBR-061).

\bibliographystyle{JHEP}
\bibliography{refs}
\end{document}